\documentclass[twocolumn]{IEEEtran} 

\usepackage[dvips]{graphicx}

\def\BibTeX{{\rm B\kern-.05em{\sc i\kern-.025em b}\kern-.08em
    T\kern-.1667em\lower.7ex\hbox{E}\kern-.125emX}}

\begin{document}

\title{DAB Content Annotation and Receiver Hardware Control with XML} 

\author{Darran Nathan, Eva Rosdiana, Chua Beng Koon\thanks{The authors are with the
DSP Technology Centre of NgeeAnn Polytechnic, Singapore. (e-mail: nrd2@np.edu.sg [Darran Nathan], rev@np.edu.sg [Eva Rosdiana], cbk@np.edu.sg [Chua Beng Koon]).}}

\markboth{arXiv.org}
{Murray and Balemi: Using the Document Class IEEEtran.cls} 

\maketitle

The Eureka-147 Digital Audio Broadcasting (DAB) standard defines the 'dynamic labels' data field for holding information about the transmission content. However, this information does not follow a well-defined structure since it is designed to carry text for direct output to displays, for human interpretation. This poses a problem when machine interpretation of DAB content information is desired. Extensible Markup Language (XML) was developed to allow for the well-defined, structured machine-to-machine exchange of data over computer networks. This article proposes a novel technique of machine-interpretable DAB content annotation and receiver hardware control, involving the utilisation of XML as metadata in the transmitted DAB frames. 


\section{Introduction}
\PARstart{T}{here} are various digital audio broadcast standards in existance. The European Eureka-147 DAB standard is described in this article. This standard offers advanced methods for source coding, channel coding, and modulation, with the DAB frames having fields defined for both data and audio. The main advantages of this standard over FM are CD-quality audio, a 'Single Frequency Network' throughout the coverage area, data transmission capability, spectrum usage efficiency, and suitability for mobile reception. 

However, one limitation of this standard is that it does not define structured data fields for annotation of the audio content. The 'dynamic-labels' field is used for carrying transmission content information, such as artiste name and song title. This field does not impose a well-defined structure on the data it holds, since it is designed to carry text that will be directly displayed to the end-user. As such, it has limited machine interpretability.

Machine interpretability of received data will allow for powerful and flexible applications at the receiver; different receivers may choose to decode different parts of the bitstream, or effect different responses to the same received data, depending on the application.

Extensible Markup Language (XML) is a specification of the World Wide Web Consortium (W3C) that allows for the description (or markup) of data using a set of well-defined tags. The utilisation of XML gives data a structure that allows for machine interpretation of that data. Its main application thus far has been in the exchange of business information over the internet.

The need for structured annotation of DAB content, and the ability of XML to provide structured annotation of data, imply a synergistic combination of both technologies in a novel application of XML and an innovative technique of DAB transmission content description - the transmission of XML over DAB.

This article begins with an introduction to the 'Campus DAB' initiative undertaken by NgeeAnn Polytechnic, which led to the innovation of 'XML over DAB' as a base technology. Next, a description of how non-audio data is transmitted in DAB frames is given. It is shown how this leads to the problem of unstructured content annotation. The following section explains how XML is designed to give data a well-defined structure, and discusses the XML schema and messaging protocol developed for both annotation of DAB transmission content and control of the receiver hardware. Next, the necessary transmission frame header settings and extensions to the existing DAB standard to allow for transmission of XML are discussed. Finally, the architectural set-up and software deployment as developed in the working prototype system are analyzed.






\section{The Campus DAB Initiative}
\label{sectCampusDAB}
Traditionally, DAB has been used by radio broadcasters who wish to offer their listeners the clarity of digital audio and the transmission of data, such as weather forecasts and traffic information, alongside that audio. The DSP Technology Centre of NgeeAnn Polytechnic has taken DAB to the next level of application, under its 'Campus DAB' initiative. This project will see DAB receivers deployed at various locations throughout the campus, each of which selectively decodes a particular subchannel of the DAB transmission, and performs an action on the decoded data depending on the receiver's location and purpose. For example, a DAB receiver located at the canteen may be configured to decode the 'Campus Radio' subchannel and output the audio through speakers, while another one located at a lift lobby may be used to decode and display announcement messages transmitted on another subchannel of the same DAB transmission ensemble. Yet another receiver located at a study area may decode e-learning content on a PC terminal.

The complication arises from the fact that different data meant for different receivers may be transmitted on the same subchannel; for example, an announcement message meant for the Mathematics Department should be decoded by a receiver located at that department, but not one located at the Engineering Department. This means that the DAB receivers need to be able to interpret the received data, and decide on a course of action depending on that data. This led to the development of the 'XML over DAB' concept that has since been implemented and tested as a prototype system at the DSP Centre. The next sections describe DAB data transmission and this prototype implementation.

\section{DAB Data Transmission}
\label{sectDataTransmission}
\begin{figure}[htb]
\begin{center}
\includegraphics[width=0.4\textwidth]{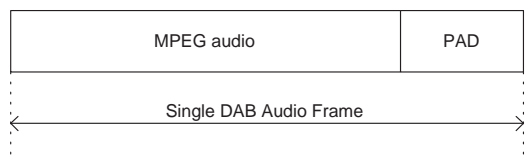}
\caption{A Single DAB Audio Frame}
\label{figDABaudioFrame}
\end{center}
\end{figure}

Data is transmitted in the Programme Associated Data (PAD) field of each DAB audio frame, as shown in Figure \ref{figDABaudioFrame}. A data object to be transmitted is split up into multiple segments, so that different segments can be sent out in the PAD fields of multiple DAB audio frames. At the receiver, these segments are extracted and combined to form the complete data object.

The two main methods of data transmission in the PAD field are Dynamic Labels and Multimedia Object Transfer (MOT)  objects \cite{bib301234}.

Dynamic Labels hold audio descriptive information for presentation to the end-user, such as song title and artiste name. The data is sent as unstructured strings of ASCII characters, for direct display by the receiver. The absence of a well-defined structure is insignificant, since the Dynamic Labels Field has been defined to carry information for human interpretation. However, a problem arises when attempting machine interpretation of the Dynamic Label. In this case, the unstructured information poses a barrier to the ability of the machine to identify portions of the Dynamic Label with their corresponding intended use. For example, the dynamic label "Dancing Queen by ABBA" can be easily understood by a human being as describing a song called "Dancing Queen" sung by "ABBA". To a machine, on the other hand, there is no way of identifying the purpose of one segment of text from another in an unstructured block. In this case, "ABBA" will have an equal probability of being a song title as "Queen" or "Dancing". This illustrates the limited use of the Dynamic Labels Field in the machine-interpretable annotation of a DAB transmission.

\begin{figure}[htb]
\begin{center}
\includegraphics[width=0.45\textwidth]{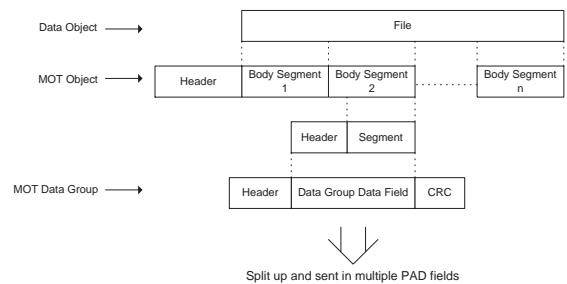}
\caption{Building the MOT Data Group}
\label{figMOTobj}
\end{center}
\end{figure}

The MOT protocol has been defined for the transmission of  data objects, such as JPEG image files. The file to be transmitted is segmented into Data Groups, as shown in Figure \ref{figMOTobj}. These Data Groups are then sent into a PAD encoder for transmission in the DAB audio frames. The MOT specification currently defines a limited, extensible set of MOT object types comprising of the ContentType and ContentSubType fields of the MOT Object header.

This explanation of the DAB non-audio data transmission capability shows that only unstructured audio annotation information can be sent in the Dynamic Labels Field. MOT objects, however, offer an avenue for transmitting data objects that can be used for providing structured audio annotation information. How this can be done is described in Section \ref{sectXMLoverDAB}.

\section{Using XML}
\label{sectXML}
The XML specification was created by the W3C to define a markup language that can be extended to suit the context of the data it describes. XML is designed to be used for the marking-up of data in a machine-to-machine information exchange environment. The mark-up is done with 'tags' that designate the properties of the enclosed data. This gives the data a well-defined structure that allows for machine interpretation of that data. However, it is the extensibility of XML that gives this markup language its true power - by utilising this capability, different sets of tags can be defined for different uses of data.

The structured data annotation capability of XML, its inherent design for machine interpretability, and the extensibility of the language, imply that an application of XML can be defined for annotation of the DAB audio bitstream. This will give the transmitted data a well-defined structure that allows machines to interpret, understand and react to audio content.

To use XML in such an environment, two main areas have to be looked into: the development of an XML schema that serves this purpose (i.e., the extension of XML), and the utilisation of a suitable XML messaging protocol.

\subsection{The XML Schema - DABml}
A schema declares the vocabulary (tags) of an XML application, as well as the usage of these tags. It was decided that the schema defined has to offer the ability to describe the transmitted bitstream content, and specify the reaction the machine should exhibit to particular content received.

As a result, the DABml schema has been designed with four main tags:

\begin{itemize}
\item $<$audioContent$>$ describes the content of the audio portion of the DAB transmission, such as the song title, artiste name and music genre;

\item $<$dataContent$>$ describes the content of the non-audio data portion of the DAB transmission, such as image files, web pages, etc;

\item $<$hardwareControl$>$ controls the DAB receiver hardware and the computer to which the receiver is attached. Such controls may include turning down the volume or recording a subchannel to the PC's harddisk;

\item $<$behaviours$>$ pulls together the descriptive $<$audioContent$>$ \& $<$dataContent$>$ tags, and the reactive $<$hardwareControl$>$ tag, to define the response to particular content received. 
\end{itemize}

\begin{figure}[htb]
\begin{center}
\includegraphics[width=0.4\textwidth]{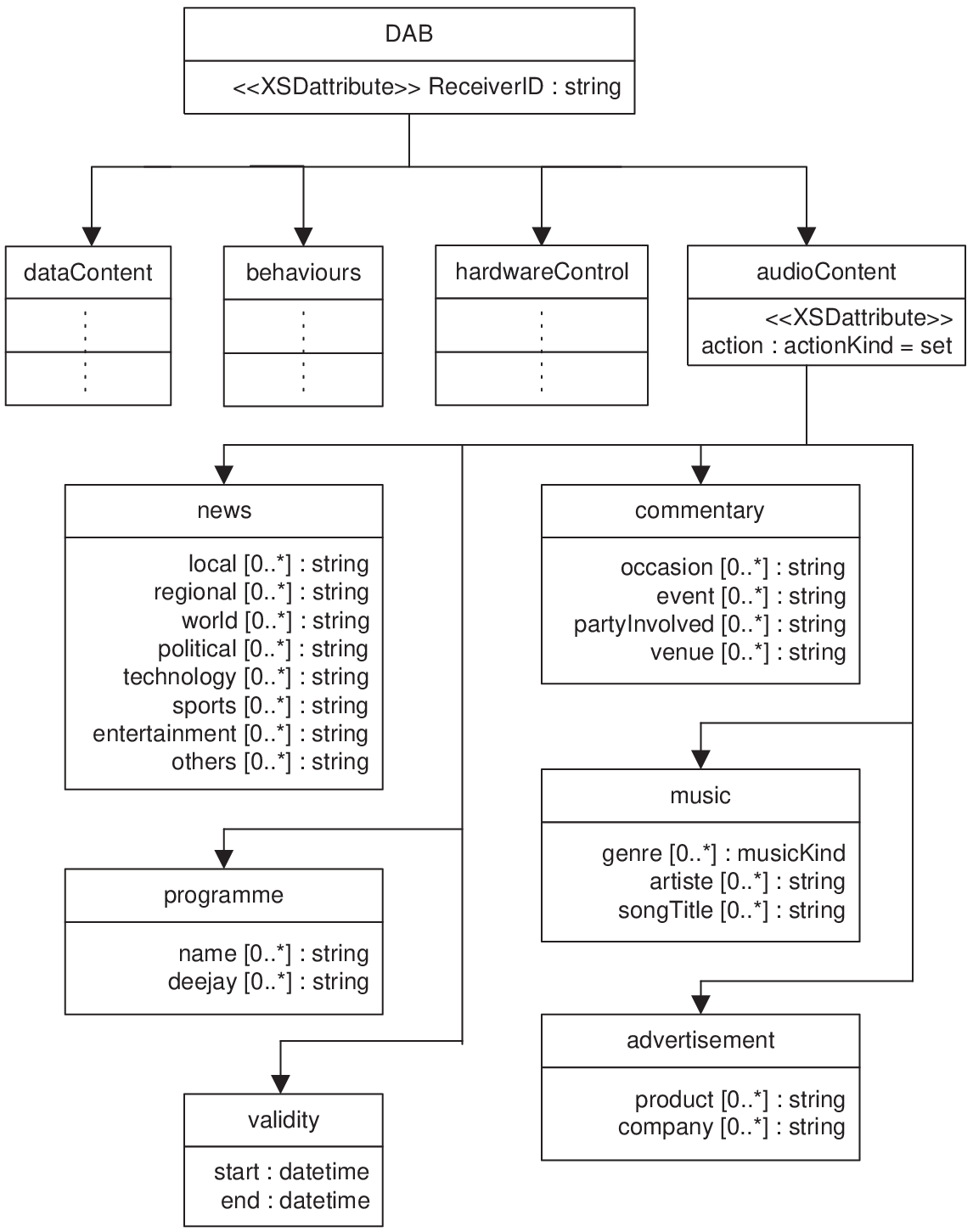}
\caption{UML Static Structure Model of the four main DABml tags}
\label{figSchemaStaticUMLmain}
\end{center}
\end{figure}

The UML (Unified Markup Language) static structure model of parts of the DABml schema is given in Figure \ref{figSchemaStaticUMLmain}.

\subsection{The XML Messaging Protocol}
To use XML in the exchange of information between a sender and a receiver, a well-defined communications protocol has to be utilised. This XML messaging protocol lays down the rules of communications between parties, so that messages are received as intended by the sender.

There are multiple XML messaging protocol specifications in existance. SOAP is one such specification that is gaining great momentum due to its open standard, widespread industry support, and simple design. As such, SOAP was selected as the DABml messaging protocol.

SOAP messages are composed of an outer envelope that encloses a header and a body. The header contains metadata about the rest of the message, while the body holds the message payload. The example below shows how SOAP can be used to transport a DABml message which specifies the artiste name and song title as 'ABBA' and 'Dancing Queen' respectively.

\begin{verbatim}

<SOAP-ENV:Envelope xmlns:SOAP-ENV= 
 "http://schemas.xmlsoap.org/soap/envelope/"
 SOAP-ENV:encodingStyle=
 "http://schemas.xmlsoap.org/soap/encoding/">

   <SOAP-ENV:Header>
   </SOAP=ENV:Header>

   <SOAP-ENV:Body>
      <dabml:DAB xmlns:dabml=
       "http://location/dabml/">
         <audioContent>
            <artiste>ABBA</artiste>
            <songTitle>Dancing Queen</songTitle>
         </audioContent>
      </dabml:DAB>
   </SOAP-ENV:Body>

</SOAP-ENV:Envelope>

\end{verbatim}

\section{Transmitting XML over DAB}
\label{sectXMLoverDAB}
The XML message to be transmitted has to be inserted into the PAD field of DAB Audio Frames, of the particular subchannel concerned. To notify the receiver that such a message is available, information is inserted into the Fast Information Channel (FIC). This requirement is not fully satisfied by the DAB standard \cite{bib300401}. However, the standard has defined several fields as being 'Reserved for future definition'. By utilising these fields to extend the current state of the DAB standard, the requirements for transmission of XML over DAB can be satisfied.

Another requirement is indication of the interval during which an XML message is 'valid'. This period of validity can, for example, indicate when the information in the XML message applies to the audio content. This allows for earlier transmission of a message that will be effective only at a specified later time.

The requirements for transmitting XML over DAB, and how these requirements are met, are detailed below:

\subsection{User Application Information}
\begin{figure}[htb]
\begin{center}
\includegraphics[width=0.3\textwidth]{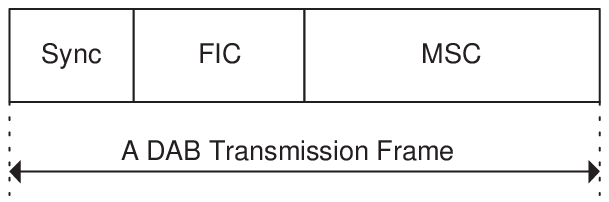}
\caption{Structure of the DAB Transmission Frame}
\label{figTransFrame}
\end{center}
\end{figure}

As shown in Figure \ref{figTransFrame}, the DAB transmission frame is divided into the FIC and the MSC (Main Service Channel). The FIC holds information about the MSC, which in turn holds all the subchannels. Information in the FIC is carried in FIGs (Fast Information Groups), each kind of which is identified by a type and extension in the format FIG TYPE/EXTENSION.

FIG 0/13 ('User Application Type') allows for signaling of the kind of data being carried in each subchannel. This is so that the user application will be able to use the correct decoder to extract the information from that subchannel. It is required that a User Application Type be defined for XML messages. The bit representation of '00000000110', defined in the DAB standard as 'Reserved for future definition', has been utilised to represent 'MOT XML Message' in this prototype implementation.

\subsection{MOT ContentType / ContentSubType}
\begin{figure}[htb]
\begin{center}
\includegraphics[width=0.4\textwidth]{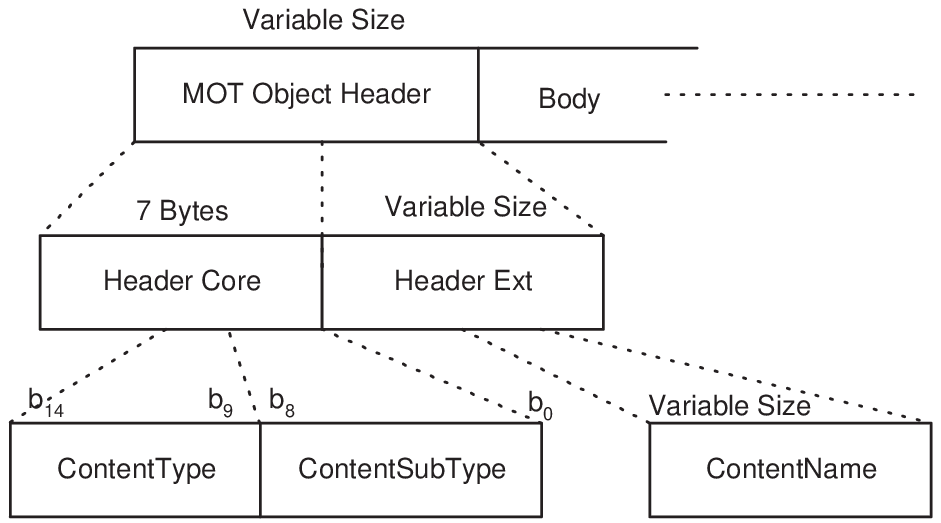}
\caption{MOT Object Structure}
\label{figContentTypeSub}
\end{center}
\end{figure}

For the MOT decoder to decide where to channel a decoded MOT object, there must be some technique of signaling the type of data in the object. The MOT specification defines the 'ContentType', 'ContentSubType' and 'ContentName' fields of the MOT object header core for this purpose, as shown in Figure \ref{figContentTypeSub}.

To transmit XML, the MOT Header 'ContentType' and 'ContentSubType' will have to contain '000000' ('General Data') and '000000001' ('MIME/HTTP') respectively, so that the Header Extension can contain the 'ContentName' string of 'TEXT/XML'. This ContentName will identify the presence of XML annotation to the receiver.

\subsection{Object Validity}
As required for the synchronized interpretation of XML messages, and as offered by the MOT specification, the validity of MOT objects is signaled by the parameter data field of the MOT object's header extension. The parameters 'StartValidity' and 'ExpireTime' are coded in UTC format, and represent the start and end of the validity period respectively.

\section{The System Architecture}
\label{sectArch}
Since SOAP is used to communicate with the DAB receiver, a computing platform is needed to extract and interpret the SOAP message before sending lower-level instructions to the DAB hardware. This computing platform may take the form of a PC which is connected to the DAB receiver via the Universal Serial Bus (USB).

If this PC is connected to the network, the possibility arises for remote communications with the DAB receiver. In this case, that PC takes on the role of a server, while any other PC on the network plays the client.

Since SOAP has been designed for client-server communications over computer network, the client in this set-up can utilise the same SOAP protocol and XML schema as that transmitted over the air, to communicate with the DAB receiver. The origin of the message destined for the receiver is transparent to the part of the server application that interprets SOAP - whether the message arrives from a client over the network or from a broadcast over the air is of concern only to the higher layers.

\subsection{Deployment of Resources}
\begin{figure}[htb]
\begin{center}
\includegraphics[width=0.5\textwidth]{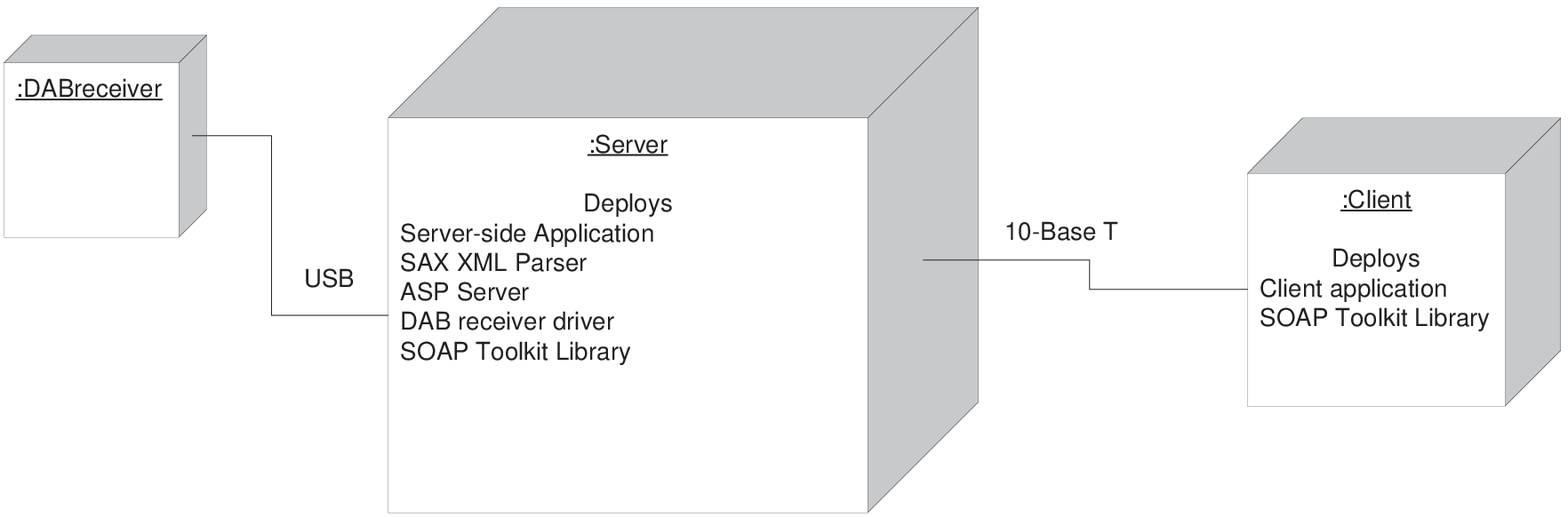}
\caption{UML Deployment Diagram of the set-up}
\label{figUMLdeployment}
\end{center}
\end{figure}

The set-up of the prototype system is illustrated in the UML Deployment Diagram of Figure \ref{figUMLdeployment}.

\subsubsection{DAB Receiver}
\begin{itemize}
\item The DAB Receiver is connected to the Server via the USB. The Server controls and receives data from the receiver over this bus.
\end{itemize}

\subsubsection{Server}
The server is the link between the DAB receiver and the 'outside world'. All DABml SOAP messages are interpreted and executed by the server. To do this, the following software are deployed:

\begin{itemize}
\item Server-side Application: This application runs various threads to extract SOAP from the subchannel bitstream, interpret the message, and perform necessary hardware control;

\item SAX XML parser: It is used by the server-side application to parse the XML in the SOAP message, so that pertinent information can be extracted;

\item ASP Server: This is the 'listener' that receives messages from network-based clients and passes them to the server-side application;

\item DAB Receiver Device Driver: The driver abstracts bus communications between the DAB receiver hardware and the server. The server-side application utilises the driver for all communications with the DAB receiver;

\item SOAP Toolkit Library: A SOAP Toolkit is a library which exposes a set of APIs to allow for transparent encoding and decoding of SOAP messages. The server-side application uses this library to send and receive SOAP messages over the network.
\end{itemize}

\subsubsection{Client}
The client communicates with the server using DABml SOAP messages. These messages may involve requests for DAB subchannel content information, or control of the hardware. The following are deployed at the client:

\begin{itemize}
\item Client Application: This software presents an interface to the end-user, which displays information about the DAB services and accepts input from the user. It acts as a 'frontend' for communications with the DAB receiver, making it transparent to the end-user that the receiver is not connected to his PC, but is instead remotely accessed over the network;

\item SOAP Toolkit Library: Just as for the case of the server, the client uses this set of APIs to encode / decode SOAP messages for communications over the network.
\end{itemize}

\subsection{How It Works}
\begin{figure}[htb]
\begin{center}
\includegraphics[width=0.5\textwidth]{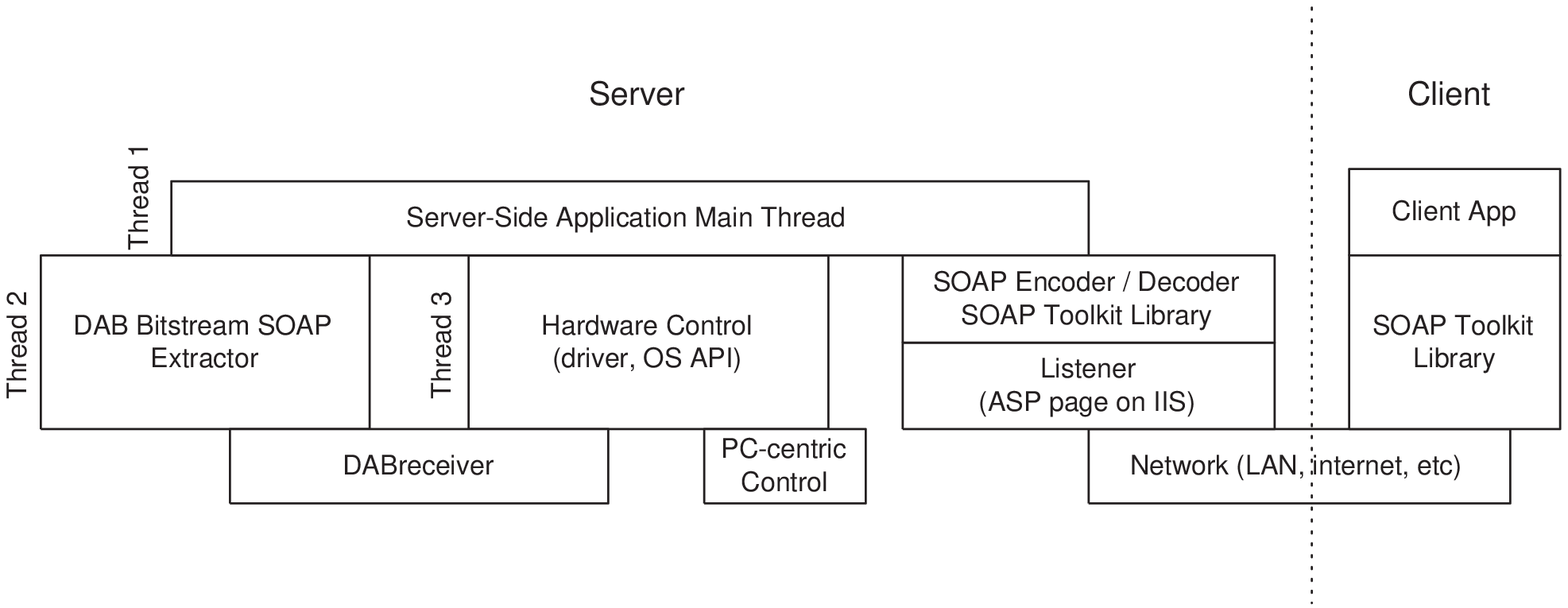}
\caption{Main Hardware and Software Layers of the Set-Up}
\label{figLayers}
\end{center}
\end{figure}

Figure \ref{figLayers} shows the main layers of software and hardware involved in the prototype set-up developed.

The server-side application consists of three threads of execution, labelled on Figure \ref{figLayers} as 'Thread 1', 'Thread 2' and 'Thread 3'.

\subsubsection{Thread 1 (Server-Side Application Main Thread) performs the following tasks:}
\begin{itemize}
\item It is the first thread started when the application is loaded. Correspondingly, it has to start the other threads;
\item SOAP messages received from either the DAB broadcast or the client are parsed using SAX, interpreted, and processed;
\item If necessary, it builds up SOAP messages for reply to the client through the SOAP Toolkit Library;
\item It monitors the set of 'behaviours' defined, and triggers a reaction whenever a message of interest is received.
\end{itemize}

\subsubsection{Thread 2 (DAB Subchannel Bitstream SOAP Extractor) performs the following tasks:}
\begin{itemize}
\item Extracts and builds up the MOT object to obtain the SOAP message;
\item Signals the Server-Side Application Main Thread and passes it the new SOAP message decoded.
\end{itemize}

\subsubsection{Thread 3 (Hardware Control) performs the following tasks:}
\begin{itemize}
\item All hardware access is done by this thread. PC resources are accessed via the Operating System (OS) API. The DAB receiver hardware is accessed via the device driver;
\item It performs all necessary initialisations of the DAB hardware upon startup, such as tuning to a default ensemble and service;
\item It engages in its own automatic monitoring and control of the DAB receiver, such as adjustment of the Automatic Frequency Control (AFC) to cater for variations in the received signal;
\item It receives hardware-control messages from the Server-Side Application Main Thread, and performs the corresponding operations on the hardware.
\item It saves decoded MOT objects from the DAB subchannel bitstream to the PC's harddisk, the URLs of which are returned to the Server-Side Application Main Thread.
\end{itemize}

\begin{figure}[htb]
\begin{center}
\includegraphics[width=0.5\textwidth]{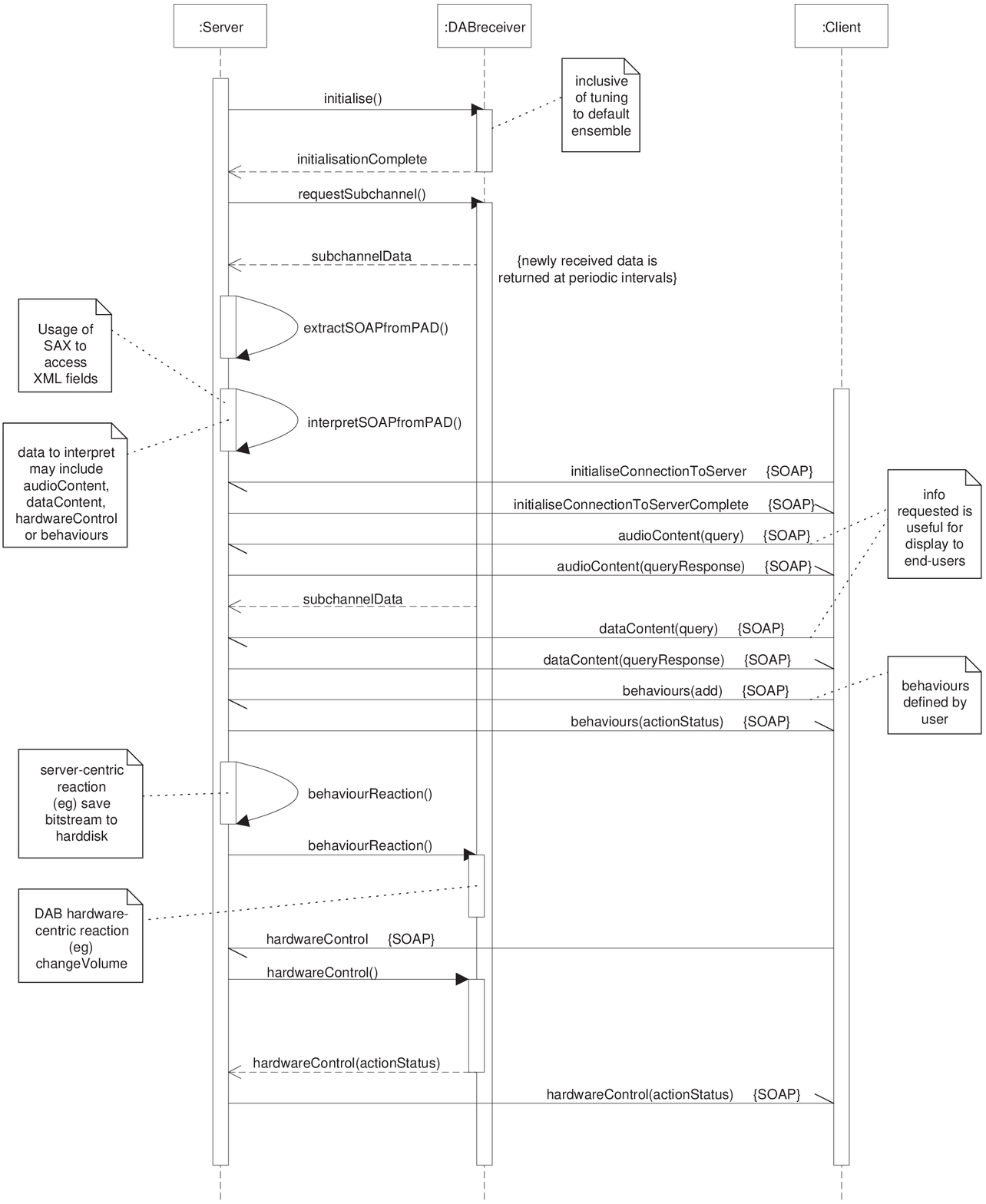}
\caption{UML Sequence Diagram of Control-Flow XML Utilisation Example}
\label{figUMLseq}
\end{center}
\end{figure}

Figure \ref{figUMLseq} shows a UML Sequence Diagram that illustrates an example of the operations that occur at the DAB receiver, the server, and the client. It also shows the calls and messages exchanged among them.

In this scenario, the server initialises the DAB receiver and requests a default subchannel. This subchannel bitstream is sent to the server at regular intervals, upon which SOAP (if present) is extracted and interpreted. Some time later, a client connects to the server and requests information about the transmitted audio and data. This information can, for example, be used as service information for display to the end-user. The client then sends a new user-defined behaviour setting to the server, to which the server responds with a message indicating success in adding the new behaviour. Next, the server invokes behaviour reactions that operate on both the server itself (server-centric reaction, such as saving a bitstream to the harddisk) and the DAB receiver (device-centric reaction, such as turning up the volume). Some time later, the client sends the server a $<$hardwareControl$>$ message (such as switching of subchannels). The server correspondingly acts on the receiver, and relays the result of the operation back to the client.

This diagram also shows that all communications between the server and the client utilise SOAP, and all communications between the client and the DAB receiver have to go through the server.

\subsection{The Prototype Implementation}
The system architecture described in this section has been implemented as a prototype at the DSP Centre, to test its feasibility for deployment under the 'Campus DAB' initiative (described in Section \ref{sectCampusDAB}).

It is planned that the 'Campus DAB' project will utilize PC-based DAB receivers, since PC resources such as harddisk space for data storage and the visual display monitor are desired at each deployment location.

The prototype has therefore been tested with both the client and server running on the same PC (to simulate a deployment location), as well as on separate PCs (to simulate a remote client being used to administer a deployment over the network). 

The prototype implementation has been shown to be ideally suited for the 'Campus DAB' initiative. In particular, the ability to remotely program the server at any deployment point from a client on any PC is a boon for administration of the Campus DAB network.

\section{Conclusions}
This article has presented a novel technique of DAB content annotation and receiver hardware control involving the use of SOAP embedded in DAB transmission frames. The transmission of PAD in DAB has been described, and the problem of unstructured data exchange explained. The features of XML that allow for well-defined, structured machine-machine exchange of data has been illustrated, and the capabilities offered by a marriage of XML and DAB explained. An XML schema (DABml), which allows for both DAB content annotation and receiver hardware control, has been proposed and implemented. The client-server architecture set-up as demonstrated in the prototype test system has been discussed and shown to be suitable for campus-wide deployment under the Centre's 'Campus DAB' initiative.

\subsection{Lessons Learned}
One trade-off of this proposal is that XML is verbose text-based metadata that can significantly increase the overall size of data to be transmitted. This results in a lowering of the efficiency of the transmission.However, the fact that XML is just text brings about the advantage that it can be sent over the network without being a cause of concern for firewalls. This allows the same markup language and protocol to be used for messages sent either over the broadcast transmission or over the computer network to PCs beyond the firewall.

A mistake made in an early design and implementation was to require transfer of XML received over the air to the client for analysis, with the 'behaviours' also stored at the clients. The rationale behind this idea was that the client-orientation would allow a single receiver to be shared among different clients, each of which picks out only the data it is interested in from the entire received DAB transmission channeled to each client.The problem encountered was that by performing subchannel selection and extraction at the client, followed by further MOT object extraction and decoding, and perhaps even decoding and output of the MPEG audio, the client PC processors were insufficiently powerful enough to handle all these tasks resulting in buffer overruns and loss of data. It was found that letting the DAB receiver hardware perform the subchannel selection and audio decoding, with the server receiving a particular subchannel for MOT data extraction / storage, and the client only involved in control and data request, resulted in an optimal distribution of processing-intensive tasks - although this meant each deployment point required a DAB receiver and server.

\subsection{Future Implementation}
Given the many incompatible digital audio broadcast standards in existance, such as IBOC (USA), DAB (Europe) and DRM, DABml can be a common metadata language used across all these standards to describe the data content of the transmission regardless of the standard. This provides for an abstraction of the actual frame structure of the transmission, allowing a common higher-level application program concerned with data interpretation and display to be used across these standards.

This can even be taken one step further to the development of a 'broadcast markup language' that will serve to describe content across all digital broadcast standards, such as DTV, DVB, XM Radio, Sirius, DAB and DRM.

\section*{Acknowledgments}
Special thanks to Oliver Faust, Bernhard Sputh, and Lim Choo Min, of NgeeAnn Polytechnic. We gratefully acknowledge the funding support provided by the NgeeAnn Kongsi (Singapore).

\nocite{*}
\bibliographystyle{IEEE}

%

%

\end{document}